\RequirePackage{fix-cm}
\documentclass[twocolumn]{svjour3}          
\smartqed  
\usepackage{graphicx}
\usepackage{multirow}
\usepackage[utf8]{inputenc}
\usepackage[english]{babel}
\usepackage[dvipsnames]{xcolor}
\definecolor{intnull2}{RGB}{214,214,214}

\usepackage{pdflscape}
\usepackage{rotating}
\usepackage{microtype}

\begin{document}

\title{Post Quantum Cryptography: Techniques, Challenges, Standardization, and Directions for Future Research
}


\author{Ritik Bavdekar \and Eashan Jayant Chopde \and Ashutosh Bhatia \and Kamlesh Tiwari,  Sandeep Joshua Daniel 
\and  Atul }

\authorrunning{Short form of author list} 

\institute{Ritik Bavdekar,  Ashutosh Bhatia, Sandeep Joshua Daniel, 
Atul \at
              Dept. of  Computer Science and Information Systems \\
Birla Insitute of Technology and Science Pilani, Pilani Campus \\ Jhunjhunu 333031, Rajasthan, INDIA \\
              \email{\{f20170349, ashutosh.bhatia, h20200135, h20200138 \}@pilani.bits-pilani.ac.in}           
          \and
          Eashan Jayant Chopde \at
              Department of Electrical and Electronics Engineering \\
Birla Insitute of Technology and Science Pilani, Pilani Campus \\ Jhunjhunu 333031, Rajasthan, INDIA \\
              \email{f20171161@pilani.bits-pilani.ac.in}
}

\maketitle

\begin{abstract}
   The development of large quantum computers will have dire consequences for cryptography. Most of the symmetric and asymmetric cryptographic algorithms are vulnerable to quantum algorithms. Grover’s search algorithm gives a square root time boost for the searching of the key in symmetric schemes like AES and 3DES. The security of asymmetric algorithms like RSA, Diffie Hellman, and ECC is based on the mathematical hardness of prime factorization and discrete logarithm. The best classical algorithms available take exponential time. Shor’s factoring algorithm can solve the problems in polynomial time. Major breakthroughs in quantum computing will render all the present-day widely used asymmetric cryptosystems insecure. This paper analyzes the vulnerability of the classical cryptosystems in the context of quantum computers, discusses various post-quantum cryptosystem families, discusses the status of the NIST post-quantum cryptography standardization process, and finally provides a couple of future research directions in this field.
\end{abstract}










\section{Introduction}
Today’s world revolves around communication. Modern society depends on Internet as a fundamental building block for any interaction between two parties. There comes a need to protect and maintain the privacy of data being transmitted. Cryptography is a field devoted to the sole purpose of data security, where countless researchers work to ensure that the privacy and integrity of data are maintained through the implementation of various security algorithms. 

A new paradigm of computing is slowly emerging in form of quantum computers which is going to drastically change the capabilities of classical computers that we are using currently. Quantum computers will be able to perform certain compuations which  are not possible by present day high performance multicore systems. The idea of quantum computer is strongly based on quantum physics i.e., understanding how things work on the sub-atomic level, initiated in the early 20th century by several great scientists like Schrödinger, Bohr, Heisenberg and Einstein, among others. Later in 1980s scientist applied quantum physics and its mathematics to model the computers that could perform certain computations at amazingly high rate than the classical computers. Quantum computers take input in the form of quantum bits and produce output. They leverages concepts of quantum mechanics for computing. Quantum computers can be used to solve problems that are not feasible to be solved by classical computers within a bounded time. Quantum computers can be used to compute results which are not possible even by present day super computers. The applications of quantum computers are countless including navigation systems, seismology, physics research, pharmaceuticals, etc. [1].  Quantum computers can be used to solve complex science problems which even present day supercomputers fail to solve. They will revolutionize artificial intelligence by giving a massive boost in computation power. The applications of quantum computers are countless. 

There are always two faces of a coin. On one hand, quantum computers provide us the hope to solve several problems across different verticals within a bounded time. On the other hand, their arrival may become a potential threat for us in certain scenarios. One such scenario is cryptanalysis. 

Cryptanalysis is the study of different techniques which can be used to find the meaning of encrypted data, without access to secret information used by the sender and receiver to encrypt and decrypt the data. Cryptanalysis is the art of code-breaking. An unfortunate fact is that the hard problems of mathematics upon which our present day cryptographic algorithms are built will become solvable on quantum computers.  This will make it easy to crack the cryptographic algorithms widely used in digital communication. The symmetric cryptographic algorithms that are widely used include Advanced Encryption Standard (AES) and 3DES (Data Encryption Statnadrd). Their bit security will be halved with the help of Grover’s algorithm [2]. Asymmetric cryptosystems like RSA, Diffie Hellman and ECC are based on the classically hard problems of Prime Factorisation and Discrete Logarithm. Quantum computers can be used to solve these computationally secure problems in no time. This severely compromises the security of the asymmetric cryptosystems.

Post Quantum Cryptography is the study of new cryptosystems which cannot be cracked by both quantum and classical computers. The cryptosystems are divided into several families based on the underlying problem upon which the security is established. These underlying problems are believed to be unsolvable by both classical and quantum computers. The major families are lattice-based cryptography, isogeny-based cryptography, non-commutative cryptography, code-based cryptography, hash-based digital signatures, and multivariate cryptography [3].

The need for better cryptographic algorithms has led to various initiatives by the NIST to create Quantum Secure algorithms. In a drive to expedite current research into the creation of a standard, entries from all across the world were received and deliberated upon to create an appropriate standard for when the need arises.

This paper discusses the implication of quantum computers on various cryptographic algorithms currently in use, by commenting on their security in the face of quantum adversaries. It also examines different widely researched algorithms (post-quantum cryptography) that would prove difficult to crack even by a quantum adversary. Finally, the survey culminates in an overview of the most viable candidates for the NIST standard and provides future research directions in the field of post-quantum cryptography.

In 2018 Mavroeidis, et al. [4] published a paper to explain how quantum computing will affect present-day cryptography. It mentioned 4 out of the 6 major post-quantum families and explained hash-based digital signatures in detail. This paper introduces and explains the major hard problems used in post-quantum cryptography for all the families. It explains each family with important simple cryptosystems. It also presents all the NIST PQC finalists. The paper ends by mentioning different future directions in which research is being done.  

\section{Cryptography}

Cryptography is the science of concealing messages using mathematics. The message is disguised using a cryptographic encryption algorithm to hide the substance of the message. The disguised data is also known as the ciphertext. The message is known as plaintext or cleartext. The ciphertext can be converted back into the message using decryption algorithms. An implementation of a cryptographic technique is called a cryptosystem. Cryptanalysis is the science of analyzing and breaking cryptosystems that are believed to be secure.

Cryptography can provide a variety of different services to users. It provides confidentiality of the data. It scrambles the data and ensures only people with the secret key (intended receivers) can unscramble and read the data. It ensures the integrity of the data. Cryptographic mechanisms exist to ensure that the data received has not been modified by a third party. It provides services for authentication. This helps the sender and receiver ensure that the other person in the communication channel is verified and that they are notcommunicating with an impostor. It helps in proving one’s identity. It also provides services for non-repudiation of the data i.e. a mechanism to prove that the sender actually sent the message.

Cryptography has many applications in the real world [5]. Computer hard disks store many business, military, and personal secrets. These can cause serious problems if they come into the wrong hands. Cryptographic techniques are used on storage devices to ensure the data inside is secure even if someone has physical access to the storage device. Cryptography also has many applications on computer networks. It is used for authentication, digital signatures of users, and time stamping important documents. Digital payment systems like Google Pay and Paytm rest on cryptographic tools to ensure the security of bank account details. Encryption and Decryption are used in emails, instant messaging tools like Whatsapp, and social media websites like Instagram and Facebook.

\subsection{Symmetric Cryptography}
Symmetric Cryptography is a branch of cryptography wherein the encryption and decryption algorithms are both executed with a single key. Given key lengths of sufficient size and randomness, it is beyond the scope of current computational resources to perform brute force attacks on such schemes. However, due to the heavy reliance on a single key, a major factor towards implementing symmetric cryptography schemes lies in the communication of the key from the sender to the receiver securely.

\subsubsection{Data Encryption Standard (DES)}
The Data Encryption Standard is a widely used cryptographic algorithm that has already been proven to be vulnerable to brute force searches in a pre-quantum world. However by increasing the complexity of the key this vulnerability may be overcome, leading to the use of modified schemes such as the Triple-DES (3DES), which may be used with 2 or 3 keys, exponentially increasing the number of operations required in an exhaustive search.

DES is based on a Feistel-Round structure, where the message is split into two halves, which undergo alternate operations in each round [6]. Each message block is 64 bits leading to 32 bits being processed per round. The first 32 bits for each round are XOR’ed with the output of a Feistel Function which takes input as the remaining 32 bits in the round. Similarly, in the next round, the unaffected bits are swapped with the processed bits, and the round is carried out again. DES algorithm  consists of 16 such rounds, started off by an Initial Permutation (IP) and ended with a Final Permutation (FP) that is the inverse of the IP. 

The Feistel Function is given below :

\begin{enumerate}
    \item \textbf{Expansion:} The 32-bit input is converted to 48-bits by duplicating and appending the first and last bits of the next and previous 4-bit block to the beginning and end of the current 4-bit block.
    
    \item \textbf{Key Mixing:} Each round corresponds to a 48-bit subkey that is obtained from the original key through a simple key schedule. The output of the Expansion stage is XOR’ed with the subkey.
    
    \item \textbf{Substitution:} In this stage, the output from the key mixing stage is split into 8 6-bit blocks which are passed through non-linear S-box transformations, the S-boxes have been predetermined by an NIST standard and can be used as a lookup table. Each S-box takes 6-bits and outputs 4-bits, the resultant output is thus converted back to 32-bits so it may be passed onto the next round.
    
    4.	\textbf{Permutation:} Finally, the S-box outputs are rearranged in a P-box in a manner that ensures that output of each S-box is distributed across 4 different S-boxes in the next round.
    
\end{enumerate}

DES can provide sufficient confusion and diffusion through the S-boxes and P-boxes to ensure the output remains random enough so the only way to break the cipher would be a brute force search. DES has already been proven to be broken and is not recommended for use.

As of today, DES has been considered as not secure. It was first broken by the DESCHALL Project in 1997, and with the increasing rise in computational power each year, the time required has been reduced to being virtually unusable as a cipher. As of 2017, the fastest DES break (for chosen-plaintext attack) was observed with the use of Rainbow Tables, obtaining the key in under 25 seconds. 

To overcome the ease of breaking DES an effort was made to increase computation trials required through increasing effective key length. This has been implemented through the 3DES Algorithm.

\subsubsection{3-DES}
Triple DES provides a solution to the vulnerability of DES due to its small (56-bit key size) by ‘expanding’ the key size using 3 different independent keys K1, K2, K3 each of 56-bits.

Encryption of the plaintext is carried out by an Encryption- Decryption-Encryption sequence. The plaintext is first encrypted using DES with key K1. The intermediate text is then decrypted using DES with key K2. Finally the result is encrypted using DES with key K3, which gives the ciphertext [6].

Based on how we choose the keys, the time complexity of Brute Force can b	e increased, by using 3 different values of K1, K2, K3 we can achieve the maximum security 3DES has to offer, with an effective key length of 168-bits.

This attempt however was also shown to be vulnerable due to the small block size increases the probability for collisions. This was exploited by the sweet32 attack [7] wherein collisions can be found in $2^{36*2}$ trials.

\subsubsection{Advanced Encryption Standard (AES)}
AES [8] is a widely used symmetric cryptographic algorithm. It relies on the randomness of ciphertext generated from multiple rounds of a well-defined, standardized  substitution-permutation network. The AES schemes in use currently employ 128-bit, 192-bit, or 256-bit keys and generally consist of 10, 12, or 14 rounds of 4 stages each. A simple key schedule is used to generate subkeys for each round. Sufficient confusion and diffusion are achieved in the ciphertext through multiple rounds of mixing and substitution.

Each Round of the AES algorithm consists of 4 stages:

\begin{enumerate}
    \item \textbf{SubByte: }In this stage, each byte is substituted with a value given by a predefined relation. These substitutions are carried out using predetermined S-boxes consisting of 256 values.
    \item \textbf{ShiftRows:} In this stage, the elements obtained from the S-boxes are then shifted right by guidelines specified in the NIST standard, or chosen independently on a need-by-need basis.
    \item\textbf{ MixColumns:} Each column is then read as a polynomial and then multiplied by an encoding polynomial. The result is then divided by an irreducible polynomial to give a polynomial in the Galois Field.
    \item \textbf{AddRoundKey:} A round key generated by a simple key schedule is added to the state array.
\end{enumerate}

All calculations are done over an Galois extension field ($2^8$). The Decryption Algorithm is the same algorithm with the same number of rounds, in reverse order.

Vulnerabilities pertaining to repeated sequences of characters in the message, improper key selection have been found and proven to be exploitable. These attacks however do not decrease the computation time required by enough an amount to make it viable for brute force searches. Bounded by current computational resources, AES can be considered ‘secure’. AES is the most widely adopted symmetric key cipher across many wireless and wired security protocols.

\subsection{Asymmetric Cryptography}

Asymmetric cryptography is the branch of cryptography where the key is divided into two different parts: a public key and a private key. The public key is given by the receiver to the sender who uses it to encrypt the data. The private key is then used by the receiver to decrypt the message. This type of cryptography ensures the authenticity and confidentiality of the message.

\subsubsection{RSA}
RSA is a public-key cryptosystem that is widely used for secure data transmission. It was invented by Ron Rivest, Adi Shamir, and Leonard Adleman in 1977.

The algorithm is based on the prime factorization of a number’s problem. The best-known solution for prime factorization in classical computers is of exponential complexity (General Number Field Sieve). The algorithm takes advantage of the above fact to encrypt data. This makes the algorithm computationally secure.

The algorithm takes advantage of one-way functions. Such functions are easy to compute but inverting them is hard (Multiplying two prime numbers is easy but factoring the number is difficult). Prime number theorem of deep number theory and the Chebhysev’s bound ensures that when randomly traversing numbers of 200 digits or more at least 1/1000 of them are prime. This makes it easy for the users of RSA to find big prime numbers of 200 digits and more. To test if a number is prime Fermat’s primality theorem is used. This gives a probabilistic result of if a number is prime or not.

Fermat’s primality test states that if a number n is prime any random number’s $(n-1)^{th}$ power is 1 in the group $Z_n$. When Fermat’s primality test is applied with more and more different random numbers and the output is 1 the probability of n being a prime number increases. If a number is not prime Fermat’s Primality Test fails for at least half the numbers in $Z_n$. Hence with every random number giving 1 as output the probability of the number n not being prime decreases by 0.5.

For secure communication, a receiver generates a private and public key. For this, the receiver searches for 2 big prime numbers at random $p,q$. The receiver then multiplies the two prime numbers to get n, which becomes part of the public key $n=p*q$. The receiver then finds a number $L=(p-1)* (q-1)$. Then a number e relatively prime to L is found.  The receiver makes ($e,n$)  pair as the public key which is used by senders to encrypt data. On the other hand, the receiver computes the private key as $d= e^{(-1)}$.This inverse is taken in the space $Z_{(p-1)(q-1)}^*$.The message is encrypted by the sender using $C’= m^e$ in the space ($Z_n$).  It is decrypted by the receiver using $m=C^d$ in the space ($Z_n$).

For large RSA key size (1024 bits and above) no efficient method for solving the problem has been found. All of the RSA cracking solutions revolve around the prime factorization of big numbers [9]. As the complexity is exponential it is impossible to crack RSA with 2048 bits key size in realistic timelines with present-day computational power. RSA is used widely in a lot of different systems that require asymmetric cryptography.

\subsubsection{Diffie Hellman Key Exchange Protocol}
Diffie Hellman Key Exchange (DHKE) protocol is based on the discrete logarithm problem. The discrete logarithm problem is based on another one-way function that is easy to compute but hard to invert. The discrete logarithm requires a big prime number N. A generator g is found for the group $Z_N$ that uniformly sweeps through all possible numbers in the group [10]. Some power p of  g in group $Z_N$ is easy to compute while finding the power p for a given  N, generator g and output o is hard. Mathematically, in the equation $o=g^p ~ mod ~ N$ given $g,p$ and $N$ it is easy to compute $o$. But, given $o,g$ and $N$ it is hard to find $p=dlog_g^o$ in group $Z_N$. Here $dlog$ stands for discrete log.

The discrete logarithm problem solution has exponential complexity. As the proof of security of Diffie Hellman Key Exchange is based on the assumption that a discrete logarithm problem is hard, cracking the Diffie Hellman Key Exchange protocol also has exponential complexity.
The Diffie Hellman Key Exchange protocol requires two shared values that are public knowledge. These include a big prime number $p$ and a generator in the space $Z_p$, $q$. Both the parties involved in the key exchange first generate a secret random number each a and b.Then they compute A and B $A = q^a ~  mod ~ p$ and $B = q^b ~ mod ~ p$.

A and B are shared as public keys. Then both the parties can use their private keys a and b  to generate a common key. $X = B^a ~ mod ~ p$ and $X = A^b ~  mod p$. Both the parties can use the common key X for further communications using symmetric algorithms.

In the Diffie Hellman key exchange protocol, there is no special authentication. A ‘Man in the Middle Attack’ can be used. The main algorithm of Diffie Hellman is based on the discrete log problem which is hard to solve. Hence the eavesdropper can act as the other party for each party successfully pulling a man in the middle attack. It can create a separate common key with both Alice and Bob. It can then act as Alice for Bob and Bob for Alice sending messages from one party to another while successfully decoding and modifying all of them.

If the parameters aren’t selected carefully a small subgroup confinement attack can be done. Cracking algorithms also take advantage of ideas like the most common prime numbers. As long as the infrastructure and supporting algorithms for Diffie Hellman are modified and made stronger it will remain as strong as the discrete logarithm problem.

\subsubsection{Elliptic Curve Cryptography (ECC)}

Elliptic Curve Cryptography is based on the elliptic curve. The elliptic curve has the property that a line can at most intersect the elliptic curve at 3 points. This idea is used for a one-way function. An operation $A \bullet A$ is defined. [11]. This is performed n times to give a result. Y is easy to compute however computing the number of times the operation was performed is very hard. $A \bullet A \bullet \dots p$ times $= Ap =O$, so $p = dlog_AO$. This is a hard mathematics problem in the group.

ECC has a max value which limits the number of points in the group. It helps in wrapping around the lines making it similar to modular arithmetic. This max value is dependent on the key size. 

ECC has gained a lot of interest in recent years as the key sizes required for security are much smaller than their RSA counter part. A 256 bit elliptic curve provides the same security as a 3072 bits RSA key. It is currently used by famous platforms like Bitcoin, Tor, and Whatsapp [12]. ECC hardness is based on the hardness of solving the discrete logarithm problem in elliptic curve groups. There are many different elliptic curves selected by NIST which are recommended for ECC systems.With ciphertext only attack the only technique to crack ECC is to find an algorithm to solve the discrete logarithm problem. No solution has yet been found to solve the problem in realistic computation times.

Side-channel attacks can be used to crack ECC. These involve making measurements on the physical implementation. They include simple timing attacks, simple power attacks, differential power attacks, and fault analysis. The variance, amplitude, and shape of voltage peaks can be analyzed by the attacker [13].

Twist-security attacks are another category of attacks on ECC [13]. These attacks are successful when several conditions are met. The users private key can be compromised under this attack. In a twist attack, the attacker carefully selects a public key and shares it with the victim to compute a shared key. They then use the shared key to reverse engineer the private key of the victim. However, they are very easy to avoid with the correct choices of curves and parameters. Standard curves suggested by NIST have been recently found out to have backdoors.

\section{Quantum Computers}
Quantum mechanics is characterized by its absurdity. It defies intuition. This is due to the concepts like superposition, entanglement, and quantum uncertainty. Quantum superposition allows a particle to be at multiple places at the same time. Quantum entanglement describes correlations between particles that are not possible in the classical world. Quantum uncertainty states that if we observe one property of a particle some other property’s information is lost [14].

The fundamental unit of information in classical computers is a bit [15]. This bit can take only two discrete values 0 and 1. In quantum information, the unit is a qubit [16]. This is a unit vector in a two-dimensional complex vector space. A qubit is represented using the ket notation.
$<\psi> = \alpha|0> + \beta|1>$
 is the notation for a qubit. Here $|0>$ and $|1>$ are basis vectors in the complex 2d space. $\alpha$ and $\beta$ are complex numbers whose sum of squares adds up to 1. For the representation of 2 or more qubits tensor product is used.
 
 A collection of quantum gates to form a circuit is called a quantum algorithm. Common quantum gates include Controlled NOT Gate, Hadamard gate, negation, and phase.

In quantum computing, measurement is an operation that takes in a qubit and returns a classical bit ( 0 or 1). If a qubit $<\psi> = \alpha|0> + \beta|1>$ is measured the probability of 0 or 1 being returned is $|\alpha|^2$ and $|\beta|^2$. Also right after a measurement, all information about the qubit is lost and it becomes classical. Post measurement the superposition is lost and depending on the value of the classical bit returned the qubit is transformed into $|0>$ state or $|1>$ state.

Quantum no-cloning property [17] states that there is no possible way to clone the state of a quantum system. In simple words, no circuit can be built which takes in as input a qubit and returns a copy along with the original qubit. Quantum entanglement is a phenomenon in which 2 or more qubits behave in a way that their quantum states can only be defined collectively. They cannot be defined separately. eg. $(|00> +|11>)/\sqrt{2}$  is a state with 2 qubits that cannot be defined individually. This kind of state is called a Bell State. If one of the qubits is measured the other qubits' quantum state is also automatically changed.

There are a variety of different physical systems that can be used as quantum devices. These include superconduction qubits, photonic qubits (polarisation of light), and others. Photonic qubits are good for long-distance communication while supercomputing qubits are better for quantum interactions.

\subsection{Quantum Key Distribution (bb84 protocol)}

The protocol is a secure method of sharing a common key between two systems. The protocol requires a quantum and a classical channel for the key distribution. The classic channel is two-way while the quantum channel can be one-way (from the sender to receiver only and no return link). The bit and basis selected for encoding and sending a qubit are both randomly selected.

The protocol considers 2 different orthogonal basis vector sets [18]. (eg. 0,90 horizontal-vertical and 45, -45 diagonal polarisation of photons). Alice and Bob decide on encodings 0 and 1 in both basis sets. Note- measuring information in one basis set makes us lose information in the other basis set. Alice generates a random key and encodes the bit using a random basis. These qubits are sent to Bob from the quantum channel. Bob uses random basis filters (diagonal or vertical/horizontal) to measure the bits. After that, both use a classical method to announce the filters they used for each bit. If they are different filters the bit is discarded. A final key is transmitted which is only known by Alice and Bob. Since Eve can’t clone the qubits and if she measures them information may be lost Bob will know when he announces a few random bits to Alice to ensure no tampering. Also, if Eve measured all the qubits using random filters she would have selected a few different filters and lost information of the qubits to measure them again.

\subsection{Quantum Algorithms}

\subsubsection{Shor's Algorithm}
Shor designed two different quantum algorithms which can crack the two major hard problems on which the security of present-day cryptosystems is based: the Prime Factorisation Problem and the Discrete Logarithm Problem. Classical solutions to these problems have exponential solutions. With the help of period finding using Quantum Fourier Transformation Shor designed these probabilistic algorithms in polynomial time. Shor’s algorithm can crack the factoring number problem in polynomial time while before this the best classical solution(General Number Field Sieve) takes exponential time [19].

Both the algorithms consist of 2 major parts. The first part is a classical part that involves converting the problem to a period-finding problem. The second part is the quantum part that takes advantage of quantum parallelism for the period-finding problem. It uses the quantum Fourier transformation algorithm to find the period.

Shor's algorithm is probabilistic. It doesn't always factor in the number in the first attempt. With the increase in the number of times, the algorithm is run the probability of getting a factor increases.

\subsubsection{Grover’s Algorithm}

It’s a quantum search algorithm that searches in an unsorted database in $O(n^{1/2})$ [2]. In classical computers, search over an unordered database is an O(n) problem. Grover’s algorithm is also probabilistic and can be repeated several times to ensure the search is complete.  To search a Grover iteration is performed. We can repeat iterations to increase the probability of finding the element in a database.

Grover’s algorithm can be used to find the mean and median of data, for finding the inverse value of a function, etc. These tools can be used by quantum cryptanalysts for cracking algorithms. It can also be used to search for keys in symmetric cryptosystems. 

\subsubsection{Simon’s Algorithm}

Simon’s algorithm is a quantum circuit that finds out if a function is 1 to 1 or 2 to 1. The property of the function states that if two input values map to the same output their XOR results in a constant b. If b is all 0s then it becomes a 1 to 1 function else it is a 2 to 1 function.

The classical solution requires $O(2^{n-1}+1)$ to find a collision. The quantum algorithm provides an exponential speedup [20].
Simon’s algorithm inspired Shor’s algorithm. Simon’s algorithm itself has applications in cryptography provided a correct query access model is available.

\subsection{Present Developments in Quantum Computers}

Quantum computing theory was introduced by Richard Feynman in 1982. Bernstein and Vazirani [21] proved that quantum computers could violate the extended Church-Turing hypothesis. They demonstrated an algorithm called the recursive Fourier transformation. Peter Shor [19] in 1994 presented algorithms that could solve certain computational problems at significantly faster speeds. Quantum computers remained largely a theoretical field until recently.

There are two different types of quantum computers- Universal quantum computers and quantum annealers. Universal quantum computers focus on qubits and quantum logic gates. They involve building machines that manipulate qubits using logical gates. The task is to conduct a wide variety of calculations using the same logical gates using quantum computer programming. Quantum annealers do not involve logical gates and are used mostly for tackling optimization problems.

Companies like D-Wave are working on quantum annealers. These computing machines are more tolerant to noise and have around 4000 qubits.  D-Wave offers its quantum annealer commercially for companies that need optimization problem solutions. Quantum annealers use the properties of qubits to find the lowest energy state of a system. This lowest energy state corresponds to the optimal solution of the optimization problem. Quantum computers are better suited for optimization problems. Unlike classical computers which have to try one combination at a time to find the optimal solution quantum computers can run many different combinations  at the same time. Industries have approached D-Wave [22] for their scheduling problems and have received optimal and quick results. A major disadvantage of quantum annealers is that they cannot be used to perform calculations like regular programming. The problem has to be modelled as an optimization problem to use the quantum annealer. In cryptography more focus is on the universal quantum computers.

Research and development of universal quantum computers is also peaking. Moore’s law [23] is beginning to approach a halt as transistors are getting smaller and smaller. There is a limit to how small transistors can get before quantum properties start interfering with the transistors. Quantum computing might be the way out to continue the growth of computing capabilities despite the physical size restriction of transistors.

Startups [23] focusing on quantum computing are on the rise. PsiQuantum, backed by Microsoft is attempting a new approach to quantum computing. The company claims it can use an optical-based approach to deliver a 1 million qubit machine which could revolutionize computing. Different startups are working on quantum computing hardware and software. 

The big technology companies are also investing a lot in quantum computer research. Google and IBM are in a race to quantum supremacy. They are working hard to build low noise hardware. Google built a 53-qubit quantum chip that outperformed classical computers in 2019. It is also investing in many startups working on quantum computing.  IBM is also investing to build quantum hardware. It has already built a few quantum computers and plans to build one with 1000 qubits by 2023 [24]. Amazon and Microsoft are partnering with startups to bring quantum computing services to AWS and Azure.

Quantum computing will cause a paradigm shift in the IT industry. Right now researchers have a variety of problems to tackle like error correction and stability of quantum computers. There is a very small pool of skilled researchers capable of working in the field. In the next few decades to come quantum computing will revolutionize every industry with new computing applications. Quantum computers are being built to improve processing power for a variety of different problems which are classically unsolvable in polynomial time. One of the first applications of quantum computers is the Shor’s algorithm. This will render the asymmetric cryptographic algorithms insecure. Hence new problems have to be found which are unsolvable even by quantum computers. This has resulted in the study of post-quantum cryptography and the NIST Post Quantum cryptography standardization process.

\section{Security of Classical Cryptography}
The advent of quantum computers has called into question the security of classical cryptographic algorithms. Symmetric cryptography’s security has been halved in bits by quantum computers. Asymmetric cryptography will turn obsolete with quantum computational resources.

\subsection{Quantum Security of DES}
DES tries to achieve sufficient randomness through multiple rounds, however, due to the comparatively smaller key size of DES (56-bits), it can be broken easily in view of current computational resources. 3-DES increases the key length, albeit at the loss of ease of encryption, to 168-bits. Increasing the keyspace to $2^168$ raises the security to be resistant to modern brute force searches. Like AES, Grover’s Algorithm can be used for Quantum Searches on the key, this reduces the number of operations required to $2^84$ which is not very secure. 3DES is also prone to collisions and therefore is increasingly susceptible to Simon's Algorithm. 3DES is also very slow and computationally taxing and cannot be considered for widespread use.

Models of SDES [25] are currently being implemented to understand the Feistel Round structure in Quantum Circuits. The implementation of the S-Box lookup tables requires many complicated gate arrangements and cannot be compromised due to its core function in providing non-linearity to the Algorithm. The complexity of implementing 3DES in Quantum Circuits may suggest some viability for use however due to the ease in breaking the core DES algorithm, it is avoided by most.

Symmetric Cryptographic Algorithms are currently considered non-implementable in a Quantum Model and are a lot more resistant to Quantum Searches compared to public-key schemes.

\subsection{Quantum Security of AES}
AES has proven to be one of the most robust cryptographic schemes currently in use, proving resilient to the level of exhaustive brute force attacks that are currently computationally viable.

The recent advancements in the development of Quantum Computers have however led people to reopen the question of AES’s security against Quantum Computer Brute Force Exhaustive Search attacks, Asymmetric Public-Key schemes such as RSA, ECC, etc have been known to be broken completely by Quantum Exhaustive Searches due to the immense parallel processing possible through superposition of qubits [26].

Various algorithms that exploit the principle of superposition have been put forward such as Simons Promise, Grover’s Search, and Shor’s Algorithm. 

Grover’s search algorithm reduces the exhaustive key search from $O(N)$ to $O((N/M)^{1/2})$ trials, where M can be reduced to 1 by choosing functions that give single solutions while implementing AES in a quantum circuit. 

A major drawback of Grover Search based cryptanalysis of symmetric cryptosystems is that the cryptosystems must be implemented as a quantum circuit.,. Many implementations of AES as a quantum circuit exist. The crack using Grover’s search algorithm requires the AES algorithm to be on a quantum circuit. This limits the application of the crack only to quantum oracles. Furthermore, as mentioned above, current algorithms are only able to reduce the trials by $N^{1/2}$, where N is the length of the key. One can simply increase the key length to preserve security, for example, moving from AES-128 to AES-256 can still be considered unbreakable in the face of Brute Force Searches.

\subsection{Quantum Security of RSA}
Shor’s factoring algorithm is a quantum circuit that can factorize big numbers in polynomial time [27]. This is a huge speedup compared to classical methods that are dependent on exponential solutions for the same. With the development of quantum computers, in the near future, the factorization problem will not be a hard problem to solve. This will make RSA cracking very easy. The first practical implementation of RSA for factorization big numbers required more than a billion qubits. This went down to 20 million qubits in late 2019 [28]. This implementation is capable of factoring 2048 RSA integers in less than 8 hours using 20 million noisy qubits. The most powerful present-day quantum computers have 50-100 qubits. In the next 25 years, the software improvements and hardware improvements will meet to make factoring prime numbers a reality.

The algorithm uses quantum Fourier transformation to take advantage of quantum parallelism. The prime number factoring problem is converted to a period finding problem. Many optimizations are done by the authors of the paper to reduce the number of qubits required for the  factorization. Ekera-Hastad’s [28] derivative of Shor’s factorization algorithm is used to decrease the number of multiplication operations. Many other optimizations like windowed arithmetic and oblivious carry runaways are used.

The main idea behind Shor’s Prime Factorisation algorithm is as follows [29].

Let $N=p*q$ such that $p$ and $q$ are prime numbers. To factorize $N$ a random number $a<N$ is selected such that $gcd(a,N)=1$. The period of $f(x)=a^x mod N$ is found(Quantum circuit). If the period r is an odd number the algorithm is repeated from the start. Now if the period is an even number we have $f(x)=f(x+r)$. The factors of N are $gcd(a^{(r/2 \pm 1)} ,n)$. The period calculation takes the help of quantum Fourier transformation.

The first part of the algorithm requires setting an input and output register where input is the superposition of all possible inputs and output is set of 0s. The input and output registers are entangled.

The period calculation takes the help of quantum Fourier transformation. Input and output registers with log2N qubits each are initialized. The input register is initialized to a superposition of all possible values from 0 to N-1. 

A quantum circuit for the function $f(x)=a^x  mod N$ is made and the input register is passed through the circuit to initialize the output register as a superposition of $f(x)$ for all $x$ in input registers. Quantum Fourier Transformation is applied to the input register. 

A measurement is then performed in the input register which gives y. Since the input and output registers are entangled this changed the quantum state of the output with $f(x_0)$.

Y/N is converted to an irreducible fraction and the value of the denominator is used as a potential value for the period of the function. If it is the period done else multiples of the denominator are tested. If they work the period is found else the registers are initialized again and the procedure is repeated(measurement value is random and may change resulting in the different output of y).

Shor’s factoring algorithm taps at the root idea behind the RSA algorithm. A replacement for the same is required in the post-quantum world. NIST is looking for new post-quantum computer era encryption algorithms.

\subsection{Quantum Security of Diffie Hellman Key Exchange Protocol}

Diffie Hellman key exchange protocol is based on the discrete logarithm problem. The idea behind Shor’s algorithm can be used to give an exponential speedup to solve the discrete logarithm problem. 

The discrete logarithm problem can be converted to a bivariate function’s period finding problem. The period finding problem can be solved using quantum Fourier transformation which takes advantage of quantum parallelism to speed up the computations.

Let $p$ and $q$ be a big prime number and a generator in $Z_p$   group respectively. $Y = q^k$. The value of $k$ has to be found to solve the problem. A bivariate function $f(x1,x2) = q^{x1}y^{x2}$  is defined. The period of the function is found using the period finding algorithm. $f(x_1+w_1,x_2+w_2)=f(x_1,x_2)$. After finding the period, we get $k=-(w_1/w_2) mod p$ [30].

The above algorithm is a polynomial solution to crack the Diffie Hellman Key Exchange Protocol. The protocol is based on the discrete logarithm problem. As the discrete logarithm problem will be computable with the development of quantum computers the protocol will become unsafe.

\subsection{Quantum Security of ECC}

ECC is based on the discrete logarithm problem in elliptic groups. Shor published two algorithms in his famous paper. One of them was for factoring big numbers and the other was a solution to the discrete logarithm problem in any group.

All possible attacks to ECC require attacking the discrete logarithm problem. There are only exponential algorithms for the same. Shor’s algorithm is a polynomial-time algorithm that takes advantage of quantum parallelism and Quantum Fourier Transformation for period finding. This can be used to solve the discrete logarithm problem and hence attack ECC in polynomial time. 

As ECC requires smaller key sizes it will be comparatively easier to crack than RSA when quantum computers are a reality. The number of qubits required to crack ECC will be significantly lesser than qubits required to crack RSA making it susceptible to attacks sooner than RSA in the future.

\section{Post Quantum Cryptography}

Post Quantum cryptography is the field of cryptography where encryption algorithms are developed which are secure from an adversary with quantum computers. All the present-day asymmetric algorithms(RSA, ECC, DH, DSA) are crackable using quantum computers. They are based on the Prime Factoring problem or the Discrete Logarithm Problem which are easy to solve on quantum computers using Shor’s Algorithm. Mathematicians and cryptographers used these number theory problems to base the security of the asymmetric algorithms. Now they have to search for new mathematical problems which can’t be solved by quantum computers easily.

Symmetric Algorithms and hash functions are comparatively secure in a post-quantum world. Grover’s Algorithm can speed up the attacks by square root complexity [31]. However, most of the algorithms can be made secure again by doubling the key size.

All the current asymmetric algorithms are based on mathematical problems for which people have searched solutions for centuries. However, the weakness they have is that quantum computers are good at parallel tasks that require one result in the end. Since the algorithms require only one result in the end a superposition of qubits can be used to parallelize all the computations and then the result can be measured. To avoid taking advantage of the parallelism of quantum computers algorithms that require several results can be used. This way parallelism of quantum computers can’t be used to its full extent.

Presently most of the post-quantum algorithms are part of 6 different families. All of them are a different family of mathematical problems which are difficult to solve even for quantum computers. These mathematical problems will form the base for the next generation of asymmetric algorithms. NIST started the Post Quantum Cryptography standardization process in 2016. It is searching for new digital signature schemes and Public Key Encryption schemes.

\subsection{Lattice-based Cryptography}

Lattice-based cryptography algorithms are based on hard mathematics lattice problems. The family of lattice problems is used in this type of cryptography. A key feature of lattice-based cryptography is that it involves security based on the worst case problems. Most of the other cryptosystems have security based on the average case [32].

Lattice is a regularly spaced grid of points stretching out to infinity. A vector is a point in this lattice ie a tuple representing the coordinates of a point. The origin is a tuple of all 0s [71]. A vector is called a far vector if it is far away from the origin and a short vector if it is close to the origin. A basis is a small set of vectors that can be used to represent the entire lattice space. These vectors can be linearly combined to represent any vector in the lattice space. For a n-dimensional lattice n vectors are selected for the basis with the condition that when a line is made with the origin and a point no other point in the set lies on the line. This way a basis can be formed. 

There are many bases for a lattice. A basis is short if it consists only of short vectors and a basis is long if it consists of long vectors. A few major hard lattice problems include-

\textbf{Short Vector Problem:} A long basis for a lattice L is given. Find a grid point in L as close to the origin as possible.

\textbf{Short Basis Problem:} A long basis for a lattice L is given. Find a short basis for the same lattice space L.

\textbf{Closest Vector Problem:} A long basis for a lattice L is given. Also, a challenge point P in the lattice space is given. Find the closest point to P in the lattice space L.

\textbf{Short Integer Solution Problem:} m vectors of n dimensions are given such that the vectors $V_i \in Z_q^n$.Find a coefficient vector $y \in Z_q^n$  such that the linear combination of y and $v_i$ gives 0. Here y are small integers like $y \in \{0,1\}^m$.

All these problems seem elementary in a small space. However, in cryptography, the lattice space has huge dimensions. The problems are also simple if a short basis is given but for a long basis, it’s a hard mathematics problem. Mathematicians as far back as the 1800s have worked on lattices. These serve as deep insights into what can and cannot be done with lattices. This gives confidence in using the lattice problems as the base for asymmetric algorithms. A major attack against them involves using lattice reduction algorithms like LLL. LLL algorithm is a polynomial algorithm used to find a relatively small basis (not smallest) for a lattice space in which a long basis has been specified.

Ajtai-Dwork [32]came up with a cryptosystem using the Shortest Vector Problem in 1997. It was cracked in 1998  by Nguyen and Ster [33]. Goldreich-Goldwasser-Halevi algorithm [34]as published based on the Closest Vector Problem. This was cracked by Nguyen in 1999 [35]. NTRU [36]was published in 1996. Over the years it has been modified and improved and the NTRU encryption system is a final candidate for the NIST Standardization process. The present NTRU is a merger of two different  second-round NTRU candidates.

\subsubsection{GGH Encryption Scheme}

A receiver’s private key is a short base and the public key is a long base. The number of dimensions in the lattice is equal to the bit size of the message. The sender uses the bad base to find a challenge point that is close to a lattice point which is the message. This message is easily decrypted by the receiver as the receiver has a short base available.

For adversaries, it is difficult to crack as they have only long bases. It is believed to be quantum-proof but was broken by a classical computer due to certain vulnerabilities. After the collection of many encrypted messages, partial information of the plaintext could be recovered.  Similar ideas with reduced vulnerabilities are used in many of the post-quantum cryptography candidates.

\subsubsection{Learning With Errors}

Learning With Errors is a subset of lattice cryptography. It takes advantage of a new trapdoor function that is easy to compute but hard to invert. $AX = B$. If A and B are given computing X is easy using Gaussian Elimination [37]. Now if random errors e are added to AX and the only information available is B and A finding X becomes a hard mathematics problem. $AX+e = B$.

This is used as the base for cryptographic algorithms. Rings Learning with Error and Module Learning with Errors are modifications of LWE that do not require big key sizes like LWE.

The equation $AX=B$ can have more equations than variables. However, equations are written so that the system is solvable. The private key is the value of all the variables. The public key is the matrices A and B. Now if a sender wants to encrypt a message the sender randomly selects a subset of the equations and adds them. In this new equation, the sender adds a big error to encode 1 and a small error to encode 0. The receiver can use the private key(value of variables) to check if the error added is big or small to decode to 1 or 0 accordingly. This problem is a hard mathematics problem making it computationally secure.

27 of the 69 algorithms submitted to NIST are lattice-based algorithms. Google implemented Learning With Errors to its google chrome browser recently.

\subsection{Hash-based Digital Signatures}

Hash-based digital signature schemes are an alternative to present day digital signature schemes which use asymmetric algorithms like RSA. They depend on 2 properties of the hash function’s collision resistance and preimage resistance.

Preimage resistance of a hash function H implies that given an output y of the hash function it is difficult to find any input x such that $y=H(x)$. Weak collision resistance of a hash function H implies that given arbitrary message m1, it is difficult to find another message m2 such that $H(m1)=H(m2)$. Strong collision resistance of a hash function H implies finding messages m1 and m2 such that $H(m1)=H(m2)$. Note: Strong collision resistance is easier to exploit because of the birthday paradox. Unlike weak collision resistance and preimage resistance which require a $O(2^{n-1})$ (where n is the number of bits of the output of the hash function) search, strong collision resistance requires $o(2^{n/2})$ search as a consequence of the birthday paradox.

Finding collisions and preimages is a difficult problem if the underlying hash function is good. Finding quantum algorithms to perform these tasks will be hard if not impossible.  Hence these hash-based digital signature schemes can be used for authentication in the post-quantum world. However, they suffer from a serious disadvantage that each digital signature can be used only once.

2 of the 69 schemes in the NIST competition are hash-based cryptosystems. Lamport introduced the hash digital signature scheme in 1979 [38] . Witernitz described a One Time Signature scheme which was significantly more efficient than Lamport’s scheme. It has the smaller key size and signature size. Merkle introduced a new scheme that combined the Witernitz approach with binary trees and called it Merkle Signature Scheme. SPHINCS+ an alternative candidate for digital signatures uses a combination of the Witernitz One-Time Signature Plus Scheme and Merkle hash trees in the Forest of Random Subsets signature scheme.

\subsubsection{Lamport Digital Signature Scheme}

The Lamport Digital Signature Scheme is a one-time signature scheme. It requires a secure hash function. For b parameter of security [4] we require a hash function that produces 2b output. Consider we require 128-bit security hence any secure hash function with 256 bits can be used.

The private key is produced using a random number generator. 256 pairs of random numbers are generated. Each number is 256 bits. This serves as the private key. Hence the size of the private key is $8b^2$.

The public key is the 512 hashes of all the random numbers generated. Hence it is also of the size $8b^2$. These values are published by the user who wants to digitally sign a document.
The signing algorithm [3] requires the hash of the message. Now as the hash of the message consists of 256 bits. For all the 256 bits(depending on the value 0 or 1) we select a number from the 256 pairs and publish it. So a sequence of 256 numbers is the digital signature. This is published along with the message.

The verification algorithm requires the verifier to hash the message. Then for each bit of the hashed message, the corresponding hash is selected from the public key.  The recipient then hashes each number of the sender’s private key ad verifies if they match with the selected public key. This scheme can only be used for one-time signature after which the private key and public key pair is discarded.

\subsection{Code-based Cryptography}
Code-based cryptography is based on error-correcting codes [3]. Computer scientists have been working on these for over 40 years. Error correction codes are codes used widely in communications to correct transmission errors. To send a message the text is sent into an error correction code. Then to the output, a few errors are randomly introduced and sent.

An example of a well-known code-based cryptosystem is the McEliece algorithm [39]. It takes the help of linear error correction codes(matrix multiplication). The receiver has a good error correction code as the private key. This is multiplied by 2 blinding matrices to produce a bad error correction code, the public key. The public key is shared with everyone. A sender sends the plaintext through the bad error correction code. Then according to the overhead, the sender adds errors. This is the final ciphertext that is transmitted. The receiver uses her good error correction code to decrypt the ciphertext.

McEliece algorithm was introduced in 1978 [40] and nobody has found a weakness in it till now. The major reason why it is not practically implemented is the size of the public key. It is much larger than its asymmetric counterparts like RSA.

In the NIST competition, 21 of the 69 algorithms are code-based cryptosystems.

\subsection{Non-Commutative Cryptography}

Non-Commutative cryptography takes the help of non-commutative groups$( A+B \neq B+A)$. A simple example is a Rubik's cube where a move sequence is a set of moves applied on the Rubik's cube [41].The addition is the concatenation of move sequences. The addition of move sequences is non-commutative. Move elimination is a move that does the opposite move. This way the previous move is eliminated and the move sequence can be written without the two moves. Move sequence negation of a move sequence consists of all the opposite moves in reverse order. This way an entire move sequence is reversed. The negation sequence of a move A is represented as -A.

\textbf{Conjugacy Problem:} 2 move sequences A and B are given. Find X such that $X+A-X=B$

Since it is non-commutative it is very hard to solve. The one-way function is easy to set up but difficult to reverse.

\subsubsection{Stickel Key Exchange Protocol }

As Diffie-Hellman key exchange protocol is not quantum-proof an alternative called Stickel Key Exchange is used. This is quantum-proof and based on non-commutative cryptography [42].

In Stickel key exchange [43] 2  sequences, A and B are defined. Also, both Alice and Bob have 2 natural numbers each which serve as their private key. Let the numbers be n and m for Alice and r and s for Bob. Alice generates a public key $PK_a=mA+ nB$ and sends it to Bob. Bob generates a public key $PK_b=rA+sB$ and sends it to Alice.Then Alice computes $K_a   = mA + PK_b+ nB = (m+r)A + (s+n)B$ and Bob computes $K_b= rA + PK_a+ sB = (r+m)A + (n+s)B$

The 2 keys are the same and cannot be computed by an adversary due to non-commutative property of the group. For an eavesdropper, the conjugacy problem has to be solved.

In the NIST competition, only 1 algorithm out of 69 is a non-commutative cryptosystem. This was broken hence no non-commutative cryptosystem will be standardized in the competition.

\subsection{Multivariate Cryptography}

Multivariate cryptography is based on the hard mathematics problem of solving a system of multivariate polynomials. Multivariate cryptosystems are all based on the multivariate quadratic map [44]. The quadratic map takes a sequence $x = (x_1,...,x_n) \in F_q^n$ and returns an output $y = (p_1 (x),...,p_m (x)) \in F_q^m$   where $p_i (x)$ are multivariate quadratic polynomials for $i = 1, .., m$ and the coefficients of the polynomials are in $F_q$ . The map is called a multivariate quadratic map P with m components and n variables.

\textbf{MQ Problem:} Given $P: F_q^n  \mapsto  F_q^m $ a multivariate quadratic map  and a target $t \in F_q^m$   find a value s such that $P(s)=t$. Here s is not unique as map P is not an injective map. 

This problem is considered a hard problem even for quantum computers. There are methods like the Grobner basis that help solve the problem. Recently many Grobner basis-like algorithms such as F4/F5 and XL [45] are used for solving the MQ Problem.

Mainly digital signature schemes are designed on top of the MQ problem. The most widely known digital signature algorithm in multivariate cryptography is the Oil and Vinegar Scheme. Rainbow is a digital signature scheme based on the Unbalanced Oil and Vinegar scheme and a finalist in the NIST competition.

\subsubsection{Oil and Vinegar Scheme}

The digital signature scheme is based on the hard mathematics MQ problem. It has a combination of Oil and Vinegar variables in the polynomials. A random quadratic map $P: F_q^n \mapsto F_q^m$  is selected. The number of variables in this map is n (variables) and the number of polynomials is m (components). The number of variables is greater than the number of polynomials. There are m oil variables and n-m vinegar variables [46].

The idea behind the scheme is that all the polynomials have quadratic terms which combined vinegar-vinegar variables and oil-vinegar variables. There are no oil-oil quadratic terms. These polynomials are dispersed using an invertible linear map to avoid differentiation between oil and vinegar variables.

If Alice wants to digitally sign a document she first makes her complete quadratic map $R=P \circ T$(combination of linear maps and quadratic map) public and keeps P and T as her private key. To digitally sign a document d she computes a hash of d using any secure hash function H. $y=H(d)$. Then she computes $s’=T^{-1}  (y)$ as T is invertible linear map. Then she finds s such that $s=P^{-1}$  (s’) and shares it. This is her signature. Anyone who receives the document can use her public key R  to compute y =R(s) and match it with the hash of the document.

For Alice computation of $s=P^{-1} (s’)$ is easy as she knows the vinegar and oil variables. She randomly selects values for vinegar variables and inserts them. As only the vinegar variables have quadratic terms she is left with a linear system of oil variables. As there are m oil variables and m equations it is easily solved using Gaussian elimination. However, due to the linear map it becomes difficult for the adversary to differentiate between oil and vinegar variables. He is stuck with an NP-hard MQ problem to solve.

The oil and Vinegar scheme [47] has an equal number of oil and vinegar variables. It was easily cracked by Kipnis and Shamir [48]. A new scheme called Unbalanced Oil and Vinegar scheme was [46] proposed where the number of vinegar variables is not equal to the number of oil variables. After a lot of deliberation, correct parameter values for n and m have been decided which are less vulnerable to attacks.

The Rainbow digital signature scheme is a scheme that has multiple UOV layers built into it [49]. This was done to make it less vulnerable to the MQ Problem attack. However, these complications gave rise to a new attack called the MinRank Problem attack.

\textbf{MinRank Problem:} k Matrices $M_i   \{i=1,2,..,k\}$ are given with n rows and m columns and a target rank r is given. The problem requires finding coefficient vector y such that the linear combination of the matrices with coefficient vector y gives rank at most r.

The increase in layers increases its vulnerability against the MinRank Problem. Optimized size of 2 layers is used in the NIST Rainbow Scheme. Beullens [50] recently came up with a new intersection attack that reduces the security of the rainbow by a few bits.

\subsection{Isogeny-based Cryptosystems}

The isogeny-based cryptosystems are based on elliptic curves. Curves are defined by the solution of a polynomial equation in 2 variables. Elliptic curves are of the form $y^2 = x^3 + ax + b$, where $4a^3   + 27b^2\neq0$. They have a property that when any two points on the elliptic curve are joined a line is formed that intersects the curve at a third point. This third point is reflected over the X-axis to find a new point that is called addition in the elliptic curve. The condition $4a^3   + 27b^2 \neq 0$ ensures no singular points are there. To double a point a tangent is drawn on the elliptic curve with that point. The tangent intersects at another point whose reflection over X-axis is taken. This gives its double. Also, scalar multiplication with n of a point P involves adding P, n times i.e.,  $[n].P=P+P+..+P (n times)$.

Elliptic Curve Diffie-Hellman(ECDH) was the first elliptic curve key exchange algorithm introduced [51]. It involved standardizing an elliptic curve in a field over a prime number p and a point P with special properties for communication. Then both Alice and Bob would select random numbers $n_a$ and $n_b$  which functioned as their private key. They calculated $P_a=[n_a].P$ and $P_b=[n_b].P$ and shared it with the other party. They then multiplied their private key to the shared point to arrive at a shared secret ie Alice calculated $[n_a].P_b= [n_a n_b].P$ and Bob calculated $[n_b].P_a= [n_a n_b].P$. This is the shared secret.
The security of this algorithm is based on the discrete logarithm problem over elliptic curves. This was theoretically cracked by Shor’s algorithm and hence is no longer safe in a world with quantum computers. After this people further studied elliptic curves and came up with isogeny-based systems.

\subsubsection{Supersingular Isogeny-based Diffie Hellman(SIDH)}

An isogeny is a homomorphic rational map between two elliptic curves $\Phi : E_0 \mapsto E_1$. In SIDH Alice and Bob generate private isogenies as their private key. They apply their isogenies to a common elliptic curve E and then share their results to the other party. The other party uses the new elliptic curve and finds another elliptic curve with the help of their private isogenies. Unlike ECDH, the final curves are not necessarily the same curve, however, they are structurally identical ie the curves are isomorphic. For elliptic curves, a number called the j-invariant can be calculated. This is the same for isomorphic curves. Hence the j-invariant is calculated individually by Alice and Bob and serves as the shared secret. An n-torsion group of an elliptic curve E[n] is a set of all points P in E that satisfy $[n].P=0$.(where 0 is the additive identity of E).

The SIDH was introduced in 2011 [52]. It involves fixing a supersingular elliptic curve E that is defined on the field $F_q$   where $q=p^2$ and $p=2^a.3^b   - 1$. Alice and Bob share the supersingular elliptic curve E along with a basis $P_a, Q_a$ for $E[2^a]$ and $P_b ,Q_b$ for $E[3^b]$(A basis is a set of points whose linear combinations can generate the entire set of points in the group).

Alice and Bob select random numbers $r_a$ and $r_b$ between 0 and $2^{a-1}$ and 0 and $3^{b-1}$ respectively. Alice calculates her isogeny  $\Phi_A$ using the kernel generated by $R_A = P_A + [r_A].Q_A$. Alice then shares $E_A=\Phi_A  (E),\Phi_A  (P_b) and \phi_A  (Q_b)$ as her public key. This helps Bob compute his second isogeny $\Psi_B$  . Bob calculates $\Psi_B$ using the kernel generated by  $\Phi_A  (P_B) + [r_B].\Phi_A  (Q_B)$. He then finds the j-invariant of $\Psi_B (E_A)$. Alice uses a similar procedure to compute her j-invariant which becomes the shared secret.

The security of the scheme is based on the $l^e$-isogeny hard problem. The $l^e$ isogeny problem is given two elliptic curves E1 and E2 and an isogeny $\Phi$ exists between the two with kernel size le, find the kernel of $\Phi$. For SIDH it involves the $2^a$-isogeny and $3^b$-isogeny hard problems. Mathematicians have worked for over 20 years but have not found any significant attacks against it for a big power e and any natural number l. A key encapsulation mechanism SIKE(Supersingular Isogeny-Based Key Encapsulation) is based on the SIDH and is part of the 3rd round alternate candidates for Key Encapsulation Mechanisms in the NIST PQC Standardization process.

\section{NIST Post Quantum Cryptography Standardization Process}

As highlighted in the previous sections, current public-key algorithms and digital signatures prove to be not secure in the face of an appropriately powerful quantum computer. The latest advancements in Quantum Computing have led to the construction of a 53-qubit Quantum Computer by Google, although this is nowhere near powerful enough to expose public-key cryptography, it has changed the question of ‘if’ they can be broken to ‘when they can be broken. Google and IBM are in a race to build noise-robust high qubits quantum computers.

As a result, many algorithms are being developed that move away from the prime factorization problem, or problems vulnerable to parallel computing. These algorithms are mainly based on Lattices and Error-Correcting Codes. Most algorithms are tending towards using hard-to-solve lattice problems due to their security in both worst-case as well as average case scenarios.

In light of the growing need for Post-Quantum Security, NIST gave out a worldwide call for submissions of problems that could be used to substitute current Public Key Encryption Schemes and Digital Signature Schemes.

69 algorithms were submitted for the NIST-  Post-Quantum cryptography competition. The best ones will be selected by 2023. These algorithms will be standardized and will be used for many years to come. This 69 consists of 20 digital signature schemes and 49 public-key encryption schemes. 26 algorithms were selected after round 1.

The algorithms in NIST round 1 were evaluated on 3 major aspects: security, cost and performance, and algorithm and implementation characteristics [53].

Security was the most important aspect. The algorithms were encouraged to be semantically secure against adaptive chosen-ciphertext attacks. NIST also considered algorithms which are semantically secure against chosen-plaintext attack. Also, digital signature schemes were required to generate unforgeable signatures concerning adaptive chosen message attack. NIST defined 5 separate security categories and included a list of other desirable security properties like side-channel resistance and resistance to multi-key attack, etc.

Cost and performance was the second most important aspect. This cost includes computational efficiency(speed of algorithm) and memory requirements(code size and RAM requirements).

Algorithm and implementation characteristics include features like simple and elegant designs, flexibility (runs on different platforms and parallelism).

After round 2 NIST has selected 15 algorithms. These algorithms can be listed out based on the type of Hard Problem used to ensure the security of such algorithms.

\subsection{Lattice Based Cryptography Candidates}

\subsubsection{CRYSTALS-KYBER (Finalist)}

CRYSTALS-KYBER is a finalist in the NIST PQC standardization process. It belongs to a family of primitives from the Cryptographic Suite for Algebraic Lattices (CRYSTAL). The same suite has been used for generating the DILITHIUM Digital Signature Algorithm. Kyber is a key encapsulation mechanism (KEM) that conforms to IND-CCA2-security. This algorithm relies on the hardness of solving the module learning-with-errors (MLWE) problem [54].

KYBER has been proposed as a public key algorithm that provides levels of security comparable to AES schemes. NIST uses AES-128,192,256 as standards for the security of level 1,3,5 respectively. Kyber flavors such as kyber-512,768,1024 can provide mildly comparable levels of security (approx. within 230 of AES-256). It is one of the most competitive proposals in place due to its performance in comparison with other proposed cryptographic schemes. The change in levels of security can be implemented by simply changing the order of the block matrices used in the algorithm.

Due to the relatively faster performance of lattice systems, and smaller public key sizes, such systems make an attractive option for implementation [55]. KYBER enforces this by displaying a key size of 800 bytes to 1.5 kilobytes, compared to the 2-kilobyte key required for present-day RSA. 

Due to its dependence on MLWE, many algebraic attacks applicable to the LWE problem must be defended against, due to public key compression, questions were raised on its effect on the cryptanalysis of the ciphertext, however, no effects have so far been found. In later versions, the compression was removed and parameters were shifted to symmetric primitives based on SHAKE256, rather than the previous SHA3-256 protocol. CRYSTALS-KYBER also shares its framework with CRYTSALS-DILITHIUM, a digital signature scheme,which makes it more attractive to implement as a complete suite. 

CRYSTALS-KYBER along with other post-quantum algorithms have been implemented by CloudFare in its Reusable Cryptographic Library. Amazon has introduced a hybrid KYBER mode for its AWS Key Management Service. Also, IBM has used the KRYSTAL Suite, KYBER and DILITHIUM in what it claims to be the world’s first quantum computing safe tape drive.

\subsubsection{SABER (Finalist)}
SABER is another lattice-based KEM algorithm which is a finalist in the NIST PQC Standardization process. It has 3 flavors namely, LightSABER, SABER, and FireSABER, offering level 1,3,5 security by the NIST requirements. The security of LightSABER, SABER, and FireSABER are equivalent to AES 128, AES 192, and AES 256 respectively. It is dependent upon the Module Learning with Rounding problem(MLWR) which differs from the MLWE problem in a sense where the errors are introduced by rounding off the values. Like CRYSTALS-KYBER, the implementation of different flavors simply requires a change in the dimensions of the block matrices used in the algorithm. Unlike KYBER, SABER uses a non-NTT method of multiplication, which is unique to this algorithm out of all the NIST Round 3 submissions.

SABER uses learning with rounding which does not require sampling from an error distribution. This decreases the pseudo-randomness of the algorithm and makes the implementation easier. Also, its security rests on only one core element. This makes it easy to implement it with applications with different security requirements.

SABER is seen to offer similar core-SVP values as CRYSTALS-KYBER, and is characterized by the simple operations used in the algorithm, making it a lot easier to implement. SABER is constant-time by design, therefore no timing faults affect the security of this scheme. It is widely projected to be used in anonymous communication.

All integer powers are modulo 2 thus reducing the algorithm bandwidth, in conjunction with LWR, makes PKE compression provably secure. At the end of Round 3 of NIST Standardization, SABER is proposed to be improved by studying side-channel attack resistance and misuse resistance. It is one of the most promising candidates for standardization.

\subsubsection{NTRU (Finalist)}
NTRU is a lattice-based finalist cryptosystem that does not derive its security from the hardness of the Ring Learning With Error(RLWE) problem or the Module Learning with Error problem. This makes it different from the other lattice-based cryptosystems. Along with Classic McEliece it is one of the oldest cryptosystems of all the submissions. Due to its age, there is extensive research and literature on attacks and problem difficulty, giving us better confidence in the algorithm. The current Round 3 submission of NTRU is a merger of Round 2 submissions NTRUEncrypt and NTRU HRSS-KEM [54].

It lacks a formal worst-case-to-average-case reduction. It has two different cost models for security analysis. The non-local model has a metric similar to the core SVP metric which is used by all the other lattice-based schemes.  It also has a local model which assigns higher security to the parameters. The non-local models offer security of category 4 at best. 

It is faster and more compact than widely used RSA. There also exist certain redundancies in the algorithm which may be removed at the cost of perfect correctness. Research into proper parameterization of NTRU has been going on since the 1990s, therefore being one of the better documented and trusted cryptosystems in Round 3. It has the advantage of not being subject to intellectual property claims, as all or most patents regarding the structures used in the cryptosystem have expired.

\subsubsection{CRYSTALS-DILITHIUM}
CRYSTALS-DILITHIUM is a digital signature algorithm that belongs to the Cryptographic Suite for Algebraic Lattices [54]. The difficulty of this algorithm relies on the Short Integer Solutions Problem and the MLWE problem.It security against chosen message attacks is based on the hardness of lattice problems over module lattices.  Its design is based on the Fiat-Shamir with Aborts technique, with the exception of sampling from a uniform distribution rather than a Gaussian distribution. This makes it much easier to implement computationally giving it an edge over its main competitor, FALCON. DILITHIUM has the smallest combination of public key and signature size compared to any lattice-based signature scheme that only uses uniform sampling [56].

DILITHIUM has the lowest core-SVP performance of all the Round 3 lattice-based cryptosystems and does not qualify for NIST Level 5 security as of now. However, DILITHIUM performs well in real-world experiments and is available for use with different parameter sets.

\subsubsection{FALCON (Finalist)}
Fast-Fourier Lattice-based Compact Signature over NTRU  is a lattice-based cryptosystem. Its security is based on the hardness of the Shortest Integer Solution problem over NTRU lattices. The framework is based on NTRU lattices, with a trapdoor sampler “Fast Fourier sampling”. It is a combination of NTRU lattices, Fast Fourier Sampling and the GPV framework (hash and sign lattice-based signature schemes)

There are security proofs of FALCON  in both the Random Oracle Model and the QROM.  FALCON requires a complex implementation with many floating-point operations, tree data structures and random sampling from discrete Gaussian distributions.

FALCON requires the smallest bandwidth amongst all the digital signature schemes. FALCON is fast in signing and verifying. It is slow in key generation. FALCON can easily be replaced into all existing protocols and applications with good overall performance. It also has a constant time implementation after round 2.

NIST selected FALCON as a finalist for the digital signature schemes [54]. It plans to standardize either FALCON or DILITHIUM. More analysis on floating-point operations causing errors and side-channel attacks have to be done. Also, more study on its sampler has to be done. FALCON’s key generation algorithm uses less than 30KB of RAM [57].

\subsection{Code-Based Cryptography Candidates}

\subsubsection{Classic McEliece}

Classic McEliece is the oldest cryptosystem present in the Round 3 of submission for the NIST PQC standard [54]. It is based on the 1979 McEliece cryptosystem which used hidden Goppa Codes. The original cryptosystem was not based on the necessity for conforming to public use computation restrictions and was able to offer one-way chosen-plaintext attack security meaning that an attacker cannot efficiently find the message from a ciphertext and public key when the message is chosen randomly [58]. The current submission modifies the original cryptosystem to provide efficient implementation and CCA security which has somewhat compromised the OW-CPA security. The current system is a merger of the NTS-KEM and Classic McEliece submissions of Round 1.

Classic McEliece is remarkable in the sense that there has been 0\% improvement in the attacks on the cryptosystem with the increase in computational resources. It utilizes the tight conversion methods of PKE to IND-CCA2 secure KEM to provide security against ROM, and on proper parameterization, QROM attacks. To protect against attacks on hashing functions, well-studied, highly unstructured hashing functions are used.

Classic McEliece is vulnerable to Information Set Decoding attacks, these can be mitigated by choosing appropriate amounts of error to ensure no unmodified text appears in the ciphertext. Side-channel resistance is obtained by ensuring constant-time implementation, i.e. giving away no information regarding runtime. Classic McEliece is resistant to CCA attacks, as it implements hashing of errors that are introduced into the ciphertext, and assuming security of hash functions, this method is comparatively slower than other attacks.

Classic McEliece produces remarkably small ciphertexts, of around 256 bytes, enabling ease of integration into network packets for transmission of data. However one of the biggest drawbacks of this cryptosystem is the large size of the public key of around 1.5 MB.

\subsubsection{BIKE (Alternative)}

BIKE is a code-based algorithm that is used for Key encapsulation. It’s the short form for Bit Flipping Key Encapsulation. It is based on the Quasi-Cyclic Moderate Density Parity Check code. It takes inspiration from the MCEliece scheme. The submitters of the algorithm changed the scheme before final submissions to round 2. They improved the bandwidth and introduced a new decoder(Black Gray Flip) [59].

The security of BIKE is based on a hard problem of coding theory, the distinguishing problem. BIKE provided parameters only for categories 1- 4 and did not provide for category 5. BIKE takes the help of assumptions on the hardness of Quasi Cyclic Syndrome Decoding and QCCF. With these assumptions, it is IND-CPA secure. It also assumes the correctness of the decoder to prove it is IND-CCA secure. Its security from attacks is based on information-set decoding and complexity estimates. However, there are side-channel attacks and CCA security concerns. This is due to their lack of confidence in their decoder which they changed recently.

As BIKE recently changed its implementation, commenting about its security on a variety of different fronts is not possible. It is however viewed by NIST as a promising candidate and has been kept in the alternate category for round 3 [54]. Once all the security concerns are addressed and worked upon it can be considered for standardization. It will serve as a trustable standard in case a fallback is found in lattice-based schemes.

\subsubsection{HQC (Alternative)}
Hamming Quasi Cyclic is a code-based public-key encryption scheme. It is based on the hardness of decisional quasi-cyclic syndrome decoding with parity problem. HQC  is IND-CPA secure and the submitters also claim that it is CCA2 secure [54].

The HQC submitters presented a new decoder that uses Reed-Muller and Reed-Solomon codes. This helped in reducing the key size by a lot. Despite the reduction in key size the public key and ciphertext are 1.6-2 time and 4-5 times the size of BIKE. The bandwidth of HQC exceeds BIKE but its key generation and decapsulation mechanisms are much faster. It uses hard-coded lookup tables for faster GF arithmetic. It also has a constant time decoding algorithm. It was susceptible to a side-channel chosen ciphertext attack. This was countered using the constant time decoding algorithm.

The major disadvantages of HQC include a non-zero decryption failure rate, large ciphertexts compared to BIKE, and larger public keys. Attacks on Hamming metrics have been studied for over 50 years and provides an assurance to security [60]. It is an efficient constant time decoding algorithm. It was selected as an alternate candidate in round 3 due to its thoroughness in security compared to other code-based algorithms. It was not selected as a finalist due to its performance characteristics.

\subsection{Multivariate Cryptography Candidates}

\subsubsection{Rainbow (Finalist)}
Rainbow is a digital signature algorithm based on a variant of the Unbalanced Oil and Vinegar (UOV) multivariate scheme. The algorithm has barely been changed since its introduction in 2005. It is characterized by its use of small signatures and an extremely fast signing and verification process. Due to the layered UOV structure, it is naturally resistant to sidechain attacks, while offering security against traditional UOV attacks such as the Kipnis-Shamir attack. However, the increase in the complexity of the structure has led to further exploitations. Since 2008 there have been no new attacks on this scheme, and it is sufficient to say even the current attacks can be protected against by choosing appropriate parameters.

Currently, Rainbow digital signature scheme is at par with NIST Level 1,3,5 and is considered to be NP-hard [54]. However, after the Round 3 submission 2 attacks have been proposed by Ward Buellens. The first reduces the security of the Rainbow 1,3,5 schemes by 7 bits, 4 bits, and 19 bits respectively, and is based on the original Kipnis-Shamir attack, this attack also exposes the UOV signature scheme. The second attack reduces the security of Rainbow 1,3,5 by 20 bits, 40 bits, and 55 bits respectively. In the light of such attacks, it is highly unlikely for NIST to consider current parameter models, and the developers must now work on establishing new parameters to maintain the level of security NIST requires. Rainbow DSA is also limited by its large public and private key sizes, reaching up to 1.8 MB.

\subsubsection{GeMSS (Alternative)}

A Great Multivariate Short Signature algorithm  is a signature scheme based on the ‘big field’ paradigm. GeMSS claims EUF-CMA security with the help of the universal unforgeability of the HFEv(Hidden Field Equations) primitive. GeMSS was inspired by another algorithm QUARTZ incorporating the latest results in multivariate cryptography to improve efficiency and security.

GeMSS is based on a well-studied mathematical problem. It supports a fast verification algorithm and the smallest signatures among all the candidates. However, it requires large public keys and long signing times. It is also difficult to implement on low-end devices. The large public keys make it difficult to implement in TLS and SSH protocols.

GeMSS submitters are trying to improve the performance of the algorithm [61]. It is competing with another promising algorithm Rainbow. Rainbow can perform well on low-end devices, has good security, and has smaller public keys. In case a major fallback occurs in the rainbow GeMSS will be the primary candidate for standardization [54]. NIST selected it as an alternative for the same reasons.

\subsection{Isogeny-based Cryptography Candidates}

\subsubsection{SIKE (Alternative)}
SIKE is the only algorithm based on Elliptic Curves. Although Elliptic Curve-based algorithms are easily broken by Shor’s Algorithm, SIKE overcomes this by using pseudorandom supersingular isogenous curves that provide one-to-one mapping of points rather than point multiplication. Due to its closeness to ECC and Diffie-Hellman protocols, if standardized it will be one of the easiest algorithms to integrate into widespread usage.
SIKE boasts of the smallest key size of all Round 3 submissions (750 bytes) even for the highest security specifications. (SIKE – Supersingular Isogeny Key Encapsulation, 2020) Due to the extent of research on Elliptic curves, parameters can be chosen and understood a lot more easier in terms of defense against sidechain attacks. However a major drawback of SIKE is the comparatively less amount of research into finding isogenies for Elliptic Curves, therefore we can not be as confident as compared to other algorithms with regards to its security. SIKE is also approximately an order of magnitude slower than the other algorithms.
SIKE is currently under debate as attacks have been found against simpler versions of the algorithms, leading to speculation with regards to the overall security of the proposed algorithm [54] The developers however have shown that with the currently submitted parameter sets, SIKE remains competitive in its security levels.

\subsection{Hash-based Digital Signature Candidates}

\subsubsection{SPHINCS+(Alternative)}
SPHINCS+ is a hash-based digital signature scheme. It is an improvement over SPHINCS ensuring multi-target attack protection. [62] Its security is proved completely based on the underlying hash function.

SPHINCS+ is based on hash-based signature schemes. These were studied before the breakthrough in public cryptosystem algorithms like RSA. It is the least likely algorithm in all the round-3 candidates to be cracked that can be broken crypt-analytically in the post-quantum world. It is a clean design with lucid specifications.  

SPHINCS+ is vulnerable to fault attacks and side-channel attacks. The submitters are working on the implementation to reduce these vulnerabilities and to improve its performance. 

The signature generation scheme is much slower than the other candidates and the signatures much larger in size. SPHINCS+ signatures with the shortest length are 4 times the size of DILITIHIUM and require 1000 times the computation power.

SPHINCS+ speed and size will require major reengineering in TLS protocols which presently use public-key encryption methods. NIST considers SPHINCS+ as a fallback in case the other algorithms fail in round 3 [54]. It will also consider standardizing it for systems that require very high security and are willing to compromise on speed and size.

\subsection{Other Cryptograhy Candidates}

\subsubsection{PICNIC}
PICNIC is a digital signature algorithm that provides security against both quantum and classical attacks. It is an algorithm that is not dependent on any number-theoretic result to proves its security. It is based on zero-knowledge proof system and symmetric-key primitives like hash functions and block ciphers [63]. The present implementation of picnic’s security is based on its hash function and the LowMC block cipher.

PICNIC has large signatures and slow signing and verifying times. It has small public keys. Also, its implementation is vulnerable to serious side-channel attacks. PICNIC is a modular design. Variants of PICNIC with AES have been explored but they require much larger key sizes.

NIST believes PICNIC is not a mature algorithm yet [54]. Its design is constantly evolving and a new concept. A big advantage is its flexibility. All of its building blocks can be chosen individually. PICNIC shows great promise and there’s a potential to increase both performance and security.  NIST selected picnic as an alternative.

\section{Performance comparison of Post Quantum cryptosystems}

Table 1 shows the level of security provided by post-quantum ciphers covered in the study. The  performance of an algorithm is normally measured by the number of clock cycles it takes for completion and the amount of circuitry/area needed to implement it on a hardware platform. It also depends on the key length or any other similar metric used within the algorithm. For this purpose, many algorithms have different versions with different parameters which affect the performance.

\begin{table}[h]
\centering
\caption{Family and Security Levels of PQC algorithms}
\label{serviceTable}
\begin{tabular}{|p{1.7cm}|p{2.9cm}|p{2.3cm}|}
\hline
\textbf{Algorithm}	& \textbf{Algorithm Family}	& \textbf{Security Level} \\
 \hline
 Classic McEliece &	Code	& 5 [64] \\
 \hline
 Saber &	Lattice	& 1, 3, 5 [64] \\
\hline
Crystals-Kyber &	Lattice	 & 1, 3, 5 [64] \\
\hline
NTRU-HRSS &	Lattice	& 1 [64] \\
\hline
NTRU-HPS &	Lattice	 & 1,3,5 [65] \\
\hline
Crystals-Dilithium & 	Lattice	 & 1, 2, 3 [64] \\
\hline
SIKE &	Isogeny &	1, 2, 3, 5 [65] \\
\hline
SPHINCS+ &	Hash &	1, 3, 5 [64] \\
\hline
\end{tabular}
\end{table}

\begin{figure*}[t]
    \centering
    \includegraphics[scale=.7]{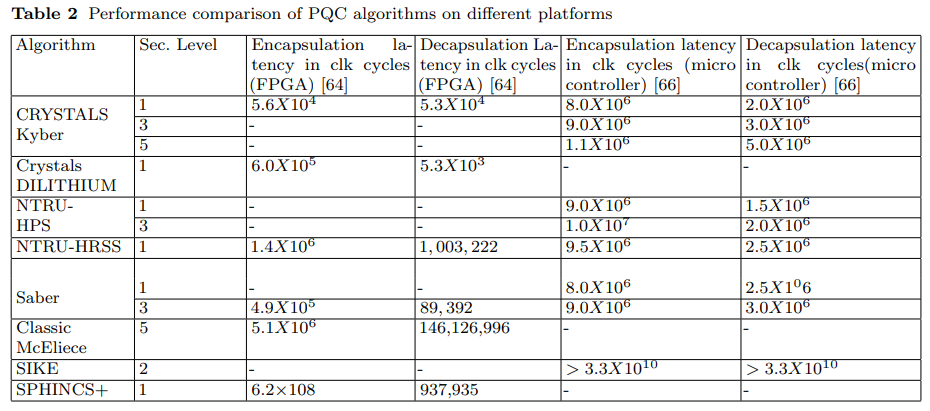}
    \label{fig:my_label}
\end{figure*}

For example, kyber512, kyber768, and kyber1024 are essentially the same algorithm but they use a different key length and other metrics which affects the performance and the security of the cipher. To get a fair comparison of performance algorithms with the same security level must be compared.


In an analysis done by Kanad Basu et al. [64], the authors found that among algorithms using key encapsulation techniques, NTRU-HRSS has the highest latency (in clock cycles). Among level 1 signature algorithms, CRYSTALS -Dilithium has lowest latency. For decapsulation techniques, NTRU-HRSS (level 3) and Classic McWliece (level 5) have the higher latency while CRYSTALS-Dilithium has least latency as shown in Table 2. These values were taken from testing the algorithms on an FPGA (Virtex 7).

From these results, we can conclude that when the base algorithms are used without any modifications, CRYSTALS-Dilithium is a good candidate for IoT for both encapsulation and decapsulation. However for decapsulation, the only practical options are level 1 algorithms for servers since all the level 5 algorithms tested do not have low latency. As IoT implementations would be more often hardware implementations testing on an FPGA serves, as a good benchmark on how the algorithm would perform on an IoT device.  The techniques like loop unrolling and loop pipelining can reduce the overall latency.

Brian Hession et al. [66] used the tended eXternal Benchmarking eXtension (XXBX) to test the performance of the PQC algorithms on a 32-bit ARM microcontroller (EK-TM4C123GXL). The algorithms were benchmarked according to the RAM, ROM, speed/latency (clock cycles) and Energy used.

In terms of RAM usage NTRU is by far the most expensive with the least usage being done by Kyber (in all 3 security levels). In terms of ROM usage, NTRU is still the most expensive, however the next most expensive algorithm is Saber, though it  uses only half the ROM space then the NTRU. When it comes to speed (clock cycles) SIKE is the slowest by a large margin, A single encapsulation takes seven minutes. Kyber, Saber and NTRU are reasonably close with higher security levels taking slightly longer time as shown in Table 2. The same pattern seen in the time consumed is present in the energy consumed. This is simply because the longer the algorithm runs the more time power is required hence the more power is consumed.

\section{Post Quantum Cryptography Migration Challenges}

The need to move to PQC is a very talked about topic, however, the issue of migration from classical cryptography  to PQC has been overlooked so far. Because of the inevitable migration to PQC algorithms, the need for research on how to migrate is very important. However, due to the  large extent of the internet and the digitalization of almost every sector, this migration is going to be more challenging than it seems. Here are some of the Gcritical challenges related to the PQC migration.

\begin{enumerate}
\item Unforeseen schedule of quantum computing development: The first risk stems from an unpredictable development schedule of QC which increases the pace of development than anticipated before.
\item Complicated Migration criteria for PQC: The migration from present-day public key cryptography to post-quantum cryptography is not as simple as upgrading to a newer version of software. The same fact has been acknowledged by NIST as well [67].
\item Record now, exploit later attacks: In this type of crypt-attack, as most of the private information is shared these days, i.e. in encrypted form, is captured and used after several years by the quantum computers if it is still relevant.

\item Selection of NIST standards relevance: It is critical to evaluate PQC migration implications during the NIST.

\end{enumerate}

There is still a lot of effort to be done to better understand migration concerns and address integration, security, performance, and other issues to get assured about safely deploying the quantum solutions in the ecosystem of global industry. Two overlapping domains should be discussed when discussing research issues:

\begin{enumerate}
    \item Research on  migration to PQC: Research on how candidate algorithms can be used to certain circumstances and regarding the safe migration for  any particular cryptographic usage domain. 
    \item 2.	Towards a Science of Cryptographic Agility:  Organizations adopt cryptographic agility as a data encryption technique to ensure a speedy response to a cryptographic attack. The goal of crypto-agility is to swiftly adapt to a new cryptographic standard without requiring large infrastructure changes. Agility in the cryptographic domain is demonstrated by the generalised extent of migration to PQC in numerous ways.
    
\end{enumerate}
There exists a lot of similarity between the large space of PQC migration and agility in the cryptographic domain. The   concerned similarities between the two signifies the difficulty range which is to be addressed to allow the migration of cryptographic domain to PQC algorithms. At the same time, there are quite a few domains where these two fields do not match or intersect at all.

\subsection{Pre PQC-Migration Challenges}

The migration from present day public key cryptography to post-quantum cryptography is not as simple as upgrading to newer version of a software which got acknowledged by NIST as well [67]. PQC addresses the key, ciphertext and signature size along with the needs in interaction and computation. Finally, NIST stated that only one algorithm would be chosen as the replacement technique because other algorithms have distinct trade-offs in terms of key size and computation requirements. Overall, NIST considers that providing many options under the new PQC standards is safe. Several areas are discussed below that could gain advantage from research:

\begin{enumerate}
    \item \textbf{Performance considerations:} As we know, PQC algorithms have higher computing, storage, memory, and communication requirements, therefore performance considerations in a variety of deployment scenarios are important.
    
    \item \textbf{Security considerations:} A lot of security issues are going to be created because of the changes in public key cryptography algorithms. As the PQC is less understood compared to the existing RSA and ECC algorithms, it can raise concerns regarding key-size, computation time, etc. Another area to be addressed is cryptanalysis of PQC algorithms.

    \item \textbf{Considerations for implementation:} PQC algorithm implementation will be more complicated than it appears. The intricacy of mathematical algorithms is reflected in this section, which translates them into platform-specific architecture and device contexts.

\end{enumerate}

Following are some of the methods that have been utilised to introduce new algorithms:

\begin{enumerate}
    \item A generally explored strategy to implement migration is to employ a hybrid strategy, in which two cryptography algorithms are utilised, one algorithm is adopted from our present set of standards and another algorithm from a recent set of PQC. This provides a mechanism to protect "record now, exploit later" while still depending on well-understood defiance to typical attacks, which can be critical for the early stages of migration. The advantage of this method is that it allows an organisation to maintain certification while transitioning to updated candidate criteria. The fact that computing, memory, connectivity, and other requirements are greatly raised is, of course, a disadvantage.

    \item Another option is to use cypher suite negotiation, which is used in IETF protocols such as TLS [68]. During the protocol's initial handshake phase, interacting parties provide a list of supported ciphersuites to choose the most robust choice that both parties support. It is possible to negotiate cypher suite version numbers, as well as further information about key sizes and parameter settings [69].  New PQC algorithms might be added as cypher suite choices using this framework, and obsolete algorithms could be deleted.  
    
    \item Formal modelling: Formal modelling of cryptographic migration is a much-needed research field. Regardless of the migration technique used, the security of the system remains a concern. Formal approaches can help to answer this question in a more fundamental way for a given scheme. This modelling is also required to assess the safety of incorporating frameworks required for migration into widely used protocols for cryptography. For instance, the TLS protocol which is extensively used might be broken down to conventional components which can resist quantum computers for research. 
    
    \item Automated tools: Without the use of automated methods, migration is almost impossible due to the size of the cryptographic infrastructure. As a result, research in this area is also required.
    
\end{enumerate}

\subsection{The Expansion of the field of Cryptographic Agility}

Cryptographic agility is an established concept in the research community, the extensiveness of PQC migration makes it necessary to expand the research area. A cryptographic agility science should be developed to encompass a greater range of  objectives, compute domains, agility modalities, and time scales. To enable cryptographic agility, frameworks and architectures should be developed and studied across a wide range of computing contexts, acceptable interfaces that meet numerous user roles.

There are few other defining challenges like inserting agility after identification of the correct area within a cryptographic solution. CA should predict the areas in which problems can arise in future and use that information to design the system.

Another challenge is the way to address cryptographic agility in old devices which can be implausible to reconfigure.  It is critical to conduct research and gain a thorough understanding of the people engaged in the process as well as the policy in place. The most important aspect to be considered is about creating the right incentives for vendors  and developers of the software products to incorporate cryptographic agility in all required products and customers and governments play an important role in this aspect. Customers are the one who can create pressure on the vendors as they are the source of business for the vendors.

\section{Future Prospects and Challenges}

The NIST PQC process will have its next meeting in June 2021. We will hear new comments on all the finalists and alternatives. Post Quantum Cryptography is a new research topic that will be ubiquitous in the coming years.

There are 6 major families upon which cryptosystems are being built. All of them have their advantages and disadvantages in terms of bit security, size and memory overhead and computational requirements  Researchers can find new mathematical problems which are hard for quantum computers and base new cryptosystems on them. They can search for new problems which require small overheads and less computational requirements.

The researchers are working hard on the NIST cryptosystem proposals. They have been cryptanalyzing the algorithms for a long time. They are searching for implementation issues, mathematical attacks, side-channel attacks , etc. This is a major research topic and cryptanalyzing the security of the finalists and alternatives will help improve the security of the algorithms and prevent standardization of algorithms that may not be secure.

The present algorithms can also be optimized by implementing new hardware and software designs. This could help decrease the time and space overhead of the algorithms. The researchers can work on techniques to decrease the computation requirements by building specialized hardware or by improving the software. They can also decrease the memory overhead.

Researchers can work on lightweight post-quantum cryptography. The increase in IoT devices have made lightweight cryptography a requirement. IoT devices communicate with each other using encrypted data as some of the data may be private(business secrets). Lightweight post-quantum cryptography is also an important field in which research has to be done. We would require algorithms that are quantum-safe and require little overhead.

Number theoretic and algorithmic research is another field where researchers can work on the post-quantum hard problems. They can work on new mathematical techniques to crack the hard problems upon which the security of the post-quantum cryptosystems is based. There is a lot of research being done on the Multivariate Quadratic map problem and the lattice based hard problems. Researchers are coming up with new solutions and techniques to solve the hard problems. 

Quantum cryptanalysis and developing new quantum algorithms is another research area that is very little explored. Shor presented two quantum algorithms that cracked the security of the present-day asymmetric algorithms. The Post Quantum Cryptography algorithms may also have different quantum algorithms which could compromise their security. Hence more research has to be done in the field.

\section{Conclusion}

Quantum Cryptography is a nascent and rapidly growing field. Many corporations around the world are pouring in resources to further the knowledge and practices we have with regards to Post Quantum Security. Due to the varied interests and studies spread out across the globe, a need to understand the current emphasis on Quantum Security and the current advancements in this field have been presented as a survey. 

Symmetric key algorithms are both classical and quantum-resistant (AES-256 has been used to characterize the highest level of security for all new algorithms), but they are difficult to implement in quantum circuits especially considering that quantum machinery has been developed only for a very small message size (approx 20 bits). Further advancements in Quantum Mechanics based technology might lead to an expansion of these capabilities, resulting in better and more efficient ways to implement symmetric cryptosystems such as AES. For symmetric cryptosystems, the quantum ways to crack the algorithm require a quantum oracle. As long as symmetric cryptography is not implemented with quantum oracles they are safe against quantum attacks. All our present-day classical data is safe. However, the implications of quantum computing on the Public Key cryptosystem are much more serious. There is no requirement for a quantum implementation of the algorithms to crack it. An adversary with local quantum resources can exploit and crack the encryption algorithms. This makes all the asymmetric enciphered data unsafe and susceptible to attacks when efficient quantum computers are built. 

As of now, Quantum algorithms already exist for all major Public Key Cryptosystems and it is only a matter of time before they are broken completely. Researchers have been trying to find ways to either increase the hardness of the problems that are currently being used (RSA, ECC) or come up with new problems that are sufficiently difficult enough even for a Quantum Computer. Of these main approaches, the latter has been more extensively utilized, with more research being encouraged into the fields of Lattice Problems, Error-Correcting Codes, Noncommutative cryptosystems, and Hash-Based Cryptosystems. Many algorithms being proposed however are difficult to implement and their performance must be optimized for widespread public use.

In 2017, NIST issued a call for algorithms around the world, in order to be able to determine a standard for Public Key Cryptography in the future. It identified that the need for defining a system was approaching fast, based on the criteria that the time for implementation along with the development of the algorithms must not exceed that of the development of systems that may break the cryptosystems currently in use. The NIST Post Quantum Cryptography competition was made up of 3 rounds of submissions, with all submissions being open to public purview, as a result, open-source contributions were able to siphon out many vulnerable submissions, with only 15 of the initial 69 making it to the final round. As of now, these 15 algorithms were analyzed and an overview has been provided. NIST has called papers for the 3rd NIST PQC Standardization Conference which will be held in June 2021. NIST plans to discuss various aspects of the algorithms and to obtain valuable feedback for informing decisions on standardization.

\section{References}
\noindent [1] T. S. Humble, “Consumer Applications of Quantum Computing: A Promising Approach for Secure Computation, Trusted Data Storage, and Efficient Applications,” IEEE Consumer Electronics Magazine, 2018. \\

\noindent [2] L. K. Grover, “A Fast Quantum Mechanical Algorithm for Database Search,” ArXiv:Quant-Ph/9605043, 1996. \\

\noindent [3] D. J. Bernstein, “Introduction to Post-Quantum Cryptography,” Springer, p. pp. 1–14, 2009. \\

\noindent [4] K. V. M. D. Z. A. J. Vasileios Mavroeidis, “The Impact of Quantum Computing on Present,” International Journal of Advanced Computer Science and Applications, 2018. \\

\noindent [5] “Cryptography in Everyday Life,” February 2021. [Online]. Available: laits.utexas.edu/~anorman/BUS.FOR/ course.mat/SSim/life.html. \\

\noindent [6] P. J. Paar C., “The Data Encryption Standard (DES) and Alternatives,” in Understanding Cryptography, Springer, 2010, pp. 55-86. \\

\noindent [7] G. Karthikeyan Bhargavan and Leurent, “On the Practical (In-)Security of 64-Bit Block Ciphers: Collision Attacks on HTTP over TLS and OpenVPN,” in Proceedings of the 2016 ACM SIGSAC Conference on Computer and Communications Security, Vienna, Austria, 2016. \\

\noindent [8] NIST, “Advanced Encryption Standard(AES),” Federal Information Processing Standards Publications, 2001. \\

\noindent [9] December 2020. [Online]. Available: https://www.sans. org/reading-room/whitepapers/vpns/paper/1006. \\

\noindent [10] P. J. Paar C., “Public-Key Cryptosystems Based on the Discrete Logarithm Problem,” in Understanding Cryptography, Springer, 2010.  \\

\noindent [11] P. J. Paar C., “Elliptic Curve Cryptosystems,” in Understanding Cryptography, Springer, 2010. \\

\noindent [12] L. Wagner, “Basic Intro To Elliptic Curve Cryptography,” December 2020. [Online]. Available: https:// qvault.io/2020/09/17/very-basic-intro-to-elliptic-curve-cryptography/. \\

\noindent [13] O. L.D., “Side-Channel Attacks in ECC: A General Technique for Varying the Parametrization of the Elliptic Curve,” in Lecture Notes in Computer Science, vol 3156, Berlin, 2004. \\

\noindent [14] M. Gilles, 2019. [Online]. Available: https://www.techn ologyreview.\\com/2019/01/29/66141/what-is-quantum-computing/.  \\

\noindent [15] V. Feldman, “Basics of Quantum Mechanics,” 2003. [Online]. Available: https://ocw.mit.edu/courses/ mathematics/ 18-435j-quantum-computation-fall-2003/lecture-notes/qc\_lec02.pdf. \\

\noindent [17] N. S. Yanofsky, “An Introduction to Quantum Computing,” ArXiv:0708.0261 [Quant-Ph], 2007. 
M. G. Kuzyk, “Quantum no-cloning theorem and entanglement,” American Journal of Physics, 2019. \\

\noindent [18] A. I. a. N. R. S. Nurhadi, “Quantum Key Distribution (QKD) Protocols: A Survey,” in 2018 4th International Conference on Wireless and Telematics (ICWT).  \\

\noindent [19] P. W. Shor, “Polynomial-Time Algorithms for Prime Factorization and Discrete Logarithms on a Quantum Computer,” SIAM Journal on Computing, p. 1484–509, 1997. \\

\noindent [20] J. Hui, “QC — Simon’s Algorithm,” April 2019. [Online]. Available: https://jonathan-hui.medium.com/qc-simons-algorithm-be570a40f6de.  \\

\noindent [21] E. a. V. U. Bernstein, “Quantum Complexity Theory,” Association for Computing Machinery, p. 11–20, 1993. \\

\noindent [22] D-WAVE, [Online]. Available: https://www.\\dwavesys.com/resources/publications?type=technology.  \\

\noindent [23] [Online]. Available: https://www.cbinsights. com/ research/report/quantum-computing/. \\

\noindent [24] IBM, [Online]. Available: https://www.ibm.com/blogs/\\research/2020/09/ibm-quantum-roadmap/. \\

\noindent [25]
M. e. a. Almazrooie, “Quantum Exhaustive Key Search with Simplified-DES as a Case Study,” SpringerPlus, 2016.  \\

\noindent [26]
X. e. a. Bogomolec, “Towards Post-Quantum Secure Symmetric Cryptography: A Mathematical Perspective,” IACR Cryptol. ePrint Arch. 2019 (2019). \\

\noindent [27]
D. J. B. a. N. H. a. P. L. a. L. Valenta, “Post Quantum RSA,” in Post-Quantum Cryptography, Springer, 2017. \\

\noindent [28]
C. a. M. E. Gidney, “How to Factor 2048 Bit RSA Integers in 8 Hours Using 20 Million Noisy Qubits,” ArXiv:1905.09749 [Quant-Ph], 2019. \\

\noindent [29]
S. Blanda, “Shor’s Algorithm – Breaking RSA Encryption,” 2014. [Online]. Available: https://blogs.ams. org/mathgradblog/2014/04/30/shors-algorithm-breaking-rsa-encryption/. \\

\noindent [30]
B. Smith, “Pre- and Post-Quantum Diffie-Hellman from Groups, Actions, and Isogenies,” ArXiv:1809.04803 [Cs], 2019.  \\

\noindent [31]
M. K. a. G. L. a. A. L. a. M. Naya-Plasencia, Breaking Symmetric Cryptosystems using Quantum Period Finding, arXiv, 2016.  \\

\noindent [32]
D. C. Ajtai M, “A public-key cryptosystem with worstcase/average-case equivalence,” in twenty-ninth annual ACM Symposium on Theory of Computing (STOC ’97), New York, 1997.  \\

\noindent [33]
P. N. a. J. Stern, “Cryptanalysis of the Ajtai-Dwork Cryptosystem,” Springer, pp. 223-242, 1998. \\

\noindent [34]
S. G. a. S. H. O. Goldreich, “Public-Key Cryptosystems from Lattice Reduction Problems,” in Advances in Cryptology -{CRYPTO} ’97, 17th Annual International Cryptology Conference, Santa Barbara, 1997. \\

\noindent [35] 
P. Nguyen, “Cryptanalysis of the Goldreich-Goldw asser- Halevi Cryptosystem,” Advances in Cryptology - CRYPTO, pp. 288-304, 1999.  \\

\noindent [36] 
J. P. a. J. H. S. J. Hoffstein, “NTRU: A ring-based public key cryptosystem,” in International Algorithmic Number Theory Symposium, 1998. \\

\noindent [37]
O. Regev, “The Learning With Errors Problem,” [Online]. Available: https://cims.nyu.edu/~regev/papers/\\lwesurvey.pdf. \\

\noindent [38]
L. Lamport, “Constructing digital signatures from a one-way function,” Palo Alto: Technical Report CSL-98 SRI International, vol. 238, 1979. \\

\noindent [39]
S. N. Biswas B., “McEliece Cryptosystem Implementation: Theory and Practice,” in PQCrypto, 2008.  \\

\noindent [40]
R. McEliece, “A public-key cryptosystem based on algebraic coding theory,” DSN Prog. Rep., Jet Prop. Lab, pp. 114-116, 1978. \\

\noindent [41]
K. Schmeh, Understanding and Explaining Post Quantum Cryptography, 2020. \\

\noindent [42]
V. Shpilrain, “Cryptanalysis of Stickel’s Key Exchange Scheme,” Springer, p. 283–88, 2008. \\

\noindent [43]
E. Stickel, “A New Method for Exchanging Secret Keys,” in Proc. of the Third International Conference on Information Technology and Applications (ICITA05) , 2005. \\

\noindent [44]
Y. B. Ding J., “Multivariate Public Key Cryptography,” in Post-Quantum Cryptography, Berlin, Springer, 2009, pp. 193-241. \\

\noindent [45] 
A. K. J. P. a. A. S. Nicolas Courtois, “Efficient algorithms for solving overdefined systems of multivariate polynomial equations,” in EUROCRYPT 2000, 2000.  \\

\noindent [46] 
P. J. G. L. Kipnis A., “Unbalanced Oil and Vinegar Signature Schemes,” in EUROCRYPT 1999, Berlin, 1999. 

\noindent [47] 
J. Patarin, “The oil and vinegar signature scheme,” in Dagstuhl Workshop on Cryptography, 1997.  \\

\noindent [48]
A. K. a. A. Shamir, “ Cryptanalysis of the oil and vinegar signature,”in CRYPTO 98, 1998.  \\

\noindent [49]
J. D. a. D. Schmidt, “Rainbow, a new multivariable polynomial signature scheme,” in International Conference on Applied Cryptography and Network, 2005 .  \\

\noindent [50]
W. Beullens, Improved cryptanalysis of UOV and Rainbow. \\

\noindent [51]
R. H. a. J. Lang, “The Elliptic Curve Diffie-Hellman (ECDH),” December 2015. [Online]. Available: http://\\koclab.cs.ucsb.edu/teaching/ecc/project/2015Projects/ Haakegaard+Lang.pdf. \\

\noindent [52]
D. F. L. Jao D., “Towards Quantum-Resistant Cryptosystems from Supersingular Elliptic Curve Isogenies,” in Post-Quantum Cryptography, Berlin, Springer, 2011, pp. 19-34. \\

\noindent [53]
L. e. a. Chen, “Report on Post-Quantum Cryptography,” NIST Internal or Interagency Report (NISTIR) 8105, National Institute of Standards and Technology, 2016. \\

\noindent [54]
G. J. A.-S. D. A. D. C. Q. D. J. K. e. a. Alagic, “Status Report on the Second Round of the NIST Post-Quantum Cryptography Standardization Process,” NIST Internal or Interagency Report (NISTIR) 8309, 2020.  \\

\noindent [55]
P. Schwabe, “Crystals Kyber,” December 2020. [Online]. Available: https://pq-crystals.org/kyber/index.shtml.

\noindent [56]
P. Schwabe, “Crystals Dilithium,” [Online]. Available: https://pq-crystals.org/dilithium/index.shtml. \\

\noindent [57] 
“Falcon,” [Online]. Available: https://falcon-sign.info/.  \\

\noindent [58] 
“Classic McEliece: Introduction,” December 2020. [Online]. Available: https://classic.mceliece.org/. \\

\noindent [59] 
“BIKE - Bit Flipping Key Encapsulation,” December 2020. [Online]. Available: https://bikesuite.org/\#spec. \\

\noindent [60] 
“HQC,” December 2020. [Online]. Available: http:// pqc-hqc.org/implementation.html. \\

\noindent [61] 
e. a. Casanova, “GeMSS- A Great Multivariate Short Signature,” December 2020. [Online]. Available: https://www-polsys.lip6.fr/Links/NIST/GeMSS.html.  \\

\noindent [62] 
P. Schwabe, “SPHINCS+,” December 2020. [Online]. Available: https://sphincs.org/.  \\

\noindent [63] 
e. a. Zaverucha, “Picnic,” December 2020. [Online]. Available: https://microsoft.github.io/Picnic/. \\

\noindent [64] 
D. S. M. N. a. R. K. Kanad Basu, NIST Post-Quantum Cryptography- A Hardware Evaluation Study, Cryptology ePrint Archive, Report 2019/047, 2019. \\

\noindent [65] 
F. F. M. A. K. M. ,. T. N. a. K. G. Viet Ba Dang, Implementation and Benchmarking of Round 2 Candidates in the NIST Post-Quantum Cryptography Standardization Process Using Hardware and Software/Hardware Co-design Approaches, Cryptology ePrint Archive, Report 2020/795, 2020.  \\

\noindent [66] 
B. a. J.-P. K. Hession, “Feasibility and Performance of PQC Algorithms on Microcontrollers,” in Second PQC Standardization Conference, NIST. \\

\noindent [67] 
NIST, Submission Requirements and Evaluation Criteria for the Post-Quantum Cryptography Standardization Process, NIST, 2017. \\

\noindent [68] 
E. Rescorla, The Transport Layer Security (TLS) Protocol Version 1.3, RFC Editor, 2018. \\

\noindent [69] 
R. Housley, Guidelines for Cryptographic Algorithm Agility and Selecting Mandatory-to-Implement Algorithms, RFC Editor, 2015.  \\

\end{document}